\documentclass[%
 amsmath,amssymb,
 aps,
 pre,
 reprint,%
]{revtex4-1}

\usepackage{graphicx}
\usepackage{dcolumn}
\usepackage{bm}
\usepackage[all]{xy}
\usepackage{subfigure}
\usepackage{sidecap}

\begin{document}


\title{Chimera states in bursting neurons}
\author{Bidesh K. Bera$^{1}$}
\email{bideshbera18@gmail.com}
\author{Dibakar Ghosh$^{1}$}
\email{diba.ghosh@gmail.com}
\author{M. Lakshmanan $^{2}$}
\email{lakshman.cnld@gmail.com}
\affiliation{$^{1}$Physics and Applied Mathematics Unit, Indian Statistical Institute, Kolkata-700108, India\\
$^{2}$Centre for Nonlinear Dynamics, School of Physics, Bharathidasan University, Tiruchirapalli-620024, India
}%

\date{\today}

\begin{abstract}
We study the existence of chimera states in pulse-coupled networks of bursting Hindmarsh-Rose neurons with nonlocal, global and local (nearest neighbor) couplings. Through a linear stability analysis, we discuss the behavior of stability function in the incoherent (i.e. disorder), coherent, chimera and multi-chimera states. Surprisingly, we find that chimera and multi-chimera states occur even using local nearest neighbor interaction in a network of identical bursting neurons alone. This is in contrast with the existence of chimera states in populations of nonlocally or globally coupled oscillators. A chemical synaptic coupling function is used which plays a key role in the emergence of chimera states in bursting neurons. Existence of chimera, multi-chimera, coherent and disordered states are confirmed by means of the recently introduced statistical measures and mean phase velocity.

\end{abstract}

\maketitle

\section{Introduction}
Synchronization in networks of neuronal systems has been an active research area due to its important role in coding and information processing in biological systems and brain. Bursting neurons are characterized by alternates of activity of neurons between a quiescent state and fast repetitive spiking on a slow time scale. There are many processes to produce bursting behavior in coupled oscillators \cite{burst}. Further, different types of synchrony occur in coupled bursting neurons which include spike synchronization, bursting synchronization, complete synchronization and anti-phase bursting synchronization \cite{burst_syn}. In general, burst synchronization occurs at lower values of coupling strength and for complete synchronization, which involves both spike and burst synchronizations, one requires a higher value of coupling strength. Two different forms of couplings are mainly used in coupled bursting neurons depending on whether the synapse is chemical \cite{chemical} or electrical \cite{electrical}. In the first case, the synaptic coupling is often approximated by a static sigmoidal nonlinear input-output function with a threshold and saturation. In the second case, the electrical coupling is a linear function and is directly dependent on the difference between the membrane potentials. Complete synchronization always occurs in globally coupled bursting neurons under suitable conditions on the coupling strength when the synaptic coupling is electrical. Using such a global synaptic coupling, identical oscillators are found to be either synchronized or oscillate incoherently, but they never exist simultaneously. Interestingly, we observe the coexistence of coherent and disordered (incoherent) states under nonlocal, global as well as local (nearest neighbor) synaptic couplings in a network of bursting neurons. Such a coexisting state was originally named as chimera state by Abrams and Strogatz \cite{strogatz} in the context of nonlocally coupled phase oscillators.

\par Chimera states in identical coupled oscillators have also been an active area  of extensive research in recent years in the field of biology, physics and social sciences \cite{strogatz, all_chimera}. Initially the chimera states have been shown to emerge when a network of identical oscillators are coupled nonlocally, that is the coupling strength decays with distance between the coupled oscillators. Chimera states are interesting because they occur even when the oscillators are identical and coupled symmetrically. There are different types of chimera states such as amplitude-mediated \cite{sen}, breathing \cite{breathing}, clustered \cite{clustered}, spiral-wave \cite{spiral}, etc. types. Chimera states have also been experimentally observed in chemical \cite{chemical_exp}, electronic \cite{electronic}, electrochemical \cite{electro} and mechanical \cite{mechanical} systems. Initially it was assumed that chimera states only exist in phase oscillators using non-local coupling configuration. But recently it has been  observed that chimera states also occur in systems exhibiting limit cycles and in chaotic dynamical systems \cite{chaos_chimera}. Very recently, chimera states have also been observed in globally coupled oscillators \cite{lakshmanan}. Chimera states observed in real-world systems \cite{real_world}, where various dynamical behaviors are involved. For example, in the case of Parkinson's disease due to lose or damage cells in the brain, synchronized activity is absent in certain regions of the brain \cite{parkinson}. In the case of epileptic seizures, specific regions of the brain are highly synchronized and the others part are not synchronized \cite{epiliptic}.

\par One of the most challenging and inspiring problems in this area is to identify the existence of chimera states in neurobiological systems. During the last decade the mechanism behind mutual synchronization and phase synchronization of chaotic bursts in neural ensembles has been explained in detail \cite{burst_syn,Igor,csps_burst}. Recently, Kalitzin et al.\cite{multistate} have observed collective dynamics of coupled neuronal oscillators which have multiple oscillatory states. In this paper we report a surprising find of chimera and multi-chimera states in networks of chaotically bursting Hindmarsh-Rose  oscillators \cite{hr_model} under different couplings, including local ones. Chimera states emerge in an ensemble of bursting neurons with at least three types of couplings, namely nonlocal, global and local (nearest neighbor) interactions. Recently, Hizanidis {\it et al.} \cite{ijbc_hr} investigated the existence of chimera states in three-dimensional Hindmarsh-Rose neuron model using a nonlocal electric type of coupling. Previously, chimera states were observed by Sakaguchi \cite{pre_HH} in coupled Hodgkin-Huxley neural oscillators with excitatory and inhibitory couplings where nonlocal coupling is essential for the appearance of these states. Very recently, Omelchenko et al.\cite{pre_robust} have discussed the robustness of chimera states in nonlocally coupled FitzHugh-Nagumo oscillators with respect to perturbations of the frequencies of the individual oscillators  and the structural transformations of the network topology.   Belykh {\it et al.} \cite{Igor} discussed the influence of coupling strength and network topology on synchronization in pulse-coupled network of bursting Hindmarsh-Rose neurons,  where no chimera like behavior has been reported.

 As noted above, chimera states have been observed in Hindmarsh-Rose neurons with electric type of nonlocal coupling \cite{ijbc_hr}, where the range of coupling and coupling strength play key roles. But the existence of chimera state in Hindmarsh-Rose neurons with chemical synaptic coupling under local, global and nonlocal interactions has not been reported earlier. In our studies a chemical synaptic coupling is used which plays a key role in the emergence of chimera states in bursting neurons for locally and globally coupled neurons. We also analytically derive the condition for mutual synchronization in globally pulse-coupled network of bursting neurons. In the case of global and local (nearest neighbor) coupling, the behavior of the stability function in incoherent, chimera and multi-chimera states are also discussed. In our analysis we have used suitable statistical measures recently introduced by Gopal et al. \cite{lakshmanan_m} as well as the notion of mean phase velocity proposed by Omelchenko et al. \cite{mpv_PRL_2013} to confirm the chimera and multi-chimera states.

\par The subsequent part of this paper is organized as follows. Sec. II is devoted to a brief presentation of  Hindmarsh-Rose model for a bursting neuron.
In Sec. III,  we numerically investigate the existence of chimera states using nonlocal coupling. In sec. IV, an analytical study of synchronization in globally coupled neurons is reported. The behavior of the stability function in chimera states is also discussed. Existence of chimera state in local (nearest neighbor) coupling is described in Sec. V.   Sec. VI provides a discussion of our results.  

\section{Hindmarsh-Rose neuron model}
\par The Hindmarsh-Rose neuron model \cite{hr_model}, which is a well known system for its chaotic behavior and different types of bursting, in its original form is expressed as follows:  
$$\dot{x}=y+ax^2-x^3-z+I,$$
$$\dot{y}=1-d x^2-y, \eqno(1)$$
$$\dot{z}=c(b(x-x_0)-z),$$
where the variable $x$ represents the membrane potential, and the variables $y$ and $z$ are the transport of ions across the membrane through the fast and slow channels, respectively. The fast variable $y$ represents the rate of change of the fast (e.g. sodium) current. The slow variable $z$ controls the speed of variation of the slow (e.g. potassium) current. This speed is controlled by the small parameter $c$. The parameter $I$ describes an external current that enters the neuron and $x_0$ is a control parameter delaying and advancing the activation of the slow current in the modeled neuron.  For the sake of simplicity, after the linear transformation / parametric redefinition \cite{Igor_trans} $x\rightarrow x, y \rightarrow 1-y, z \rightarrow  1+I+z, d\rightarrow a+\alpha, e\rightarrow -1-I-b x_0$, Eq.(1) can be written in the form 
$$\dot{x}=ax^2-x^3-y-z,$$
$$ \dot{y}=(a+\alpha)x^2-y, \eqno{(2)}$$
$$ \dot{z}=c(bx-z+e).$$
This transformed model (2) is a phenomenological model that can exhibit all common dynamical features found in a number of biophysical modelling studies of bursting. We consider $c$ a small positive parameter so that $z(t)$ varies much slower than $x(t)$ and $y(t)$. Square-wave bursting has been observed for the following set of parameter values: $a=2.8, \alpha=1.6, c=0.001, b=9$ and $e=5$ \cite{Igor}. The system (2) is monostable, that is coexistence of stable equilibrium point and limit cycle has not been observed for this set of parameter values. 

\section{Non-local interaction}
\par Now we consider a network of identical Hindmarsh-Rose neurons with non-local interaction as follows:
$$\dot{x_i}=ax_{i}^{2}-x_{i}^{3}-y_{i}-z_{i}+\frac{k}{2p}(v_{s}-x_{i})\sum_{j=i-p}^{j=i+p} c_{ij} \Gamma (x_{j}),$$
$$\dot{y_i}=(a+\alpha )x_{i}^{2}-y_{i}, \hskip 150pt \eqno{(3)}$$
$$\dot{z_i}=c(bx_{i}-z_{i}+e),\;\;\;i=1, 2, \cdot  \cdot  \cdot  , N \hskip 120pt$$  
where $N$ is the total number of elements in the network, $p$ is the number of coupled nearest neighbors in each direction on a ring so that the coupling radius $r=\frac{p}{N}$ and $k$ is the synaptic coupling strength.  The connectivity matrix  $C=(c_{ij})_{n \times n}$ is such that $c_{ij}=1$ if the $i^{th}$ neuron is connected to the $j^{th}$ neuron, otherwise $c_{ij}=0$ and $c_{ii}=0.$ The synapse is excitatory for the reversal potential $v_s=2>x_i(t)$ for all times $t$ and $x_i(t)$. The synaptic coupling function $\Gamma (x)$ is modeled by the sigmoidal nonlinear input-output function as 
$$\Gamma (x)=\frac{1}{1+e^{-\lambda (x- \Theta_{s})}}, \eqno{(4)}$$
where $\lambda$ determines the slope of the function and $\Theta_s$ is the synaptic threshold. This often-used coupling form has been called the fast threshold modulation \cite{chemical}. We choose the threshold $\Theta_s=-0.25$ so that every spike in the isolated neuron burst can reach the threshold. We fixed the value of $\lambda=10$ throughout the work. From a physicist's perspective, equation (3) represents a network of $N$ identical pulse-couped oscillators with nonlocal attractive interaction. But from a neuroscientist's point of view, such a network corresponds to an interaction between $N$-nonlocally coupled excitatory neurons with direct excitatory synapses \cite{ren}.
\begin{figure}[ht]
  \centering
  \includegraphics[width=0.52\textwidth]{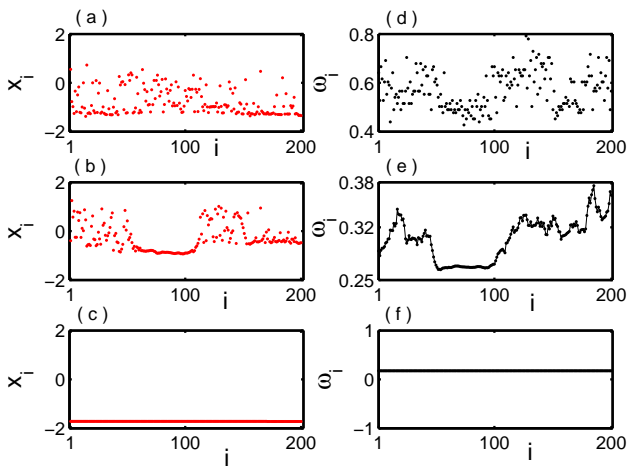} \\[-0.4cm]
  \caption{(Color online) Left panels show snapshots of amplitudes for (a) disordered state at $k=0.3$, (b) chimera state at $k=0.85$ and (c) coherent state at $k=1.4$. Right panels (e), (f) and (g) show the corresponding mean phase velocity $\omega_{i}, i=1, 2, \cdot \cdot \cdot , N,$ for each neuron. The coupling radius is fixed at $r=0.3,$ where $N=200.$}
  \label{Fig1}
\end{figure}

\par Now, we consider numerically the existence of chimera states and transitions of chimera and multi-chimera states as well as disordered and coherent states  in the network of nonlocally coupled  Hindmarsh-Rose oscillators. The network (3) and their variants in the following sections are integrated using the fifth order Runge-Kutta-Fehlberg integration algorithm scheme with integration step length $\Delta t=0.01$.  The initial conditions for (3) are chosen as follows: $x_{i0}=0.01(i-\frac{N}{2}), y_{i0}=0.02(i-\frac{N}{2}), z_{i0}=0.03(i-\frac{N}{2})$ for $i=1, 2, \cdot \cdot \cdot , \frac{N}{2}$ and for rest of the oscillators $x_{i0}=0.1(\frac{N}{2}-i), y_{i0}=0.12(\frac{N}{2}-i), z_{i0}=0.21(\frac{N}{2}-i)$ with added small random fluctuations.  In Fig. 1, we fix the coupling radius as $r=0.3$ and vary the synaptic coupling strength $k$. In the left panel, the snapshots of the amplitude (membrane potential) for disordered, chimera, and coherent states for $k=0.3, 0.85,$ and $1.4$, respectively, are shown. To confirm chimera or coherent states in nonlocally coupled neurons, first we calculate the mean phase velocity recently proposed by Omelchenko et al. \cite{mpv_PRL_2013} of each neuron as $\omega_i=2 \pi M_i/ \Delta T, i=1, 2, 3, \cdot \cdot \cdot , N,$ where $M_i$ is the number of bursts of the $i$th neuron during a sufficiently long time interval $\Delta T.$ Figure 1 (right panel) shows the mean phase velocity of each neuron corresponding to incoherent, chimera and coherent states for the synaptic coupling strength $k=0.3, 0.85$ and $1.4$ respectively. To calculate the mean phase velocity, the time interval is taken over $4\times 10^5$ time units after an initial transient of $1\times 10^5$ units. 
\par To clearly distinguish the disordered, chimera, multi-chimera and coherent states, we also use the recently introduced statistical measures by Gopal et al. \cite{lakshmanan_m} using the time series of the network. For this purpose we will calculate the strength of incoherence (SI) and discontinuity measure (DM) from a local standard deviation analysis. To calculate these statistical measures,  we first introduce a transformations $w_{1, i}=x_i - x_{i+1}, w_{2, i}=y_i - y_{i+1}, w_{3, i}=z_i - z_{i+1},\;\; i=1, 2, \cdot \cdot \cdot , N$ [see Ref. \cite{lakshmanan_m} for more details]. Different synchronization states in the network can be quantified by using the standard deviations given by
$$\sigma_l= \left \langle \sqrt{\frac{1}{N}\sum_{i=1}^N [w_{l, i}-\left \langle w_l \right \rangle]^2}~~\right\rangle_t, \eqno{(5)}$$
where $l=1, 2, 3; i=1, 2, \cdot \cdot \cdot , N,$ and $\left \langle w_l \right \rangle=\frac{1}{N}\sum_{i=1}^N w_{l, i}(t)$ and $\left \langle \cdot  \cdot  \cdot \right \rangle_t$ denotes the average over time. For coherent states the values of $\sigma_l$'s are zero while they take non-zero values for both disordered and chimera states. To distinguish chimera and disordered states, we divide the number of oscillators into $M$ (even) bins of equal length $n=N/M.$ Then we introduce local standard deviation which is defined as 
$$\sigma_l(m)=\left \langle \sqrt{\frac{1}{n}\sum_{j=n(m-1)+1}^{mn}[w_{l, j}-\left \langle w_l \right \rangle]^2}~~\right \rangle_t, \eqno{(6)}$$
$m=1, 2, \cdot \cdot \cdot ,M.$ The above quantity $\sigma_l(m)$ can be calculated for every successive $n$ number of oscillators. Then the strength of incoherence (SI) is defined as 
$$\mbox{SI}=1-\frac{\sum_{m=1}^Ms_m}{M},\;\;s_m=\Theta(\delta-\sigma_l(m)), \eqno{(7)}$$ 
where $\Theta(\cdot )$ is the Heaviside step function and $\delta$ is a predefined threshold. Consequently, the values of SI=1 or SI=0 or $0<\mbox{SI}<1$ represent disordered, coherent and chimera or multi-chimera states respectively. Again in order to distinguish chimera and multi-chimera states we also introduce the so called discontinuity measure (DM) \cite{lakshmanan_m} which is defined as  
$$\mbox{DM}=\frac{\sum_{i=1}^M \mid s_{i+1}-s_i\mid}{2}, \;\;\;\; \mbox{with} \;\; s_{M+1}=s_1.  \eqno{(8)}$$ 
For chimera state the value of $\mbox{DM}=1$ and for multi-chimera state the value of $\mbox{DM}$ is a positive integer greater than ``1".
\begin{figure}[ht]
  \centering
  \includegraphics[width=0.5\textwidth]{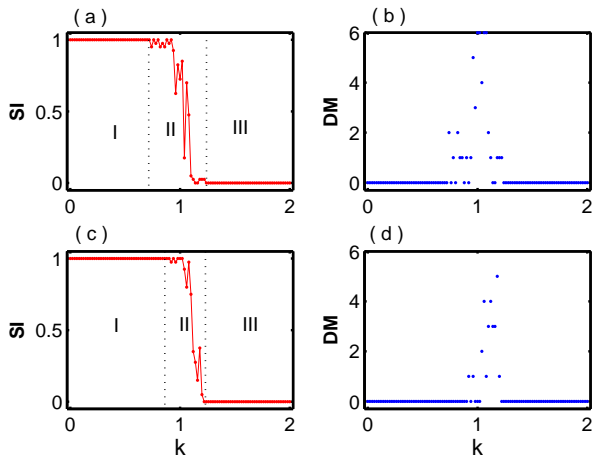} \\[0.6cm]
  \caption{(Color online) Strength of incoherence SI and discontinuity measure DM are plotted against synaptic coupling strength $k$ for two values of coupling radius $r$. (a, b) and (c, d) correspond to $r=0.3$ and $r=0.4$, respectively. Here $N=200, M=40$ and $\delta=0.05.$ }
  \label{Fig1}
\end{figure}

\par Fig. 2 shows the variation of the strength of incoherence and discontinuity measure for fixed value of coupling radius $r$ and different values of synaptic coupling strength $k$. To start with we choose $N = 200$ and the total number of bins to be $M = 40$ which we find to be optimal. For $r=0.3$, the variation of SI (Fig. 2(a)) and DM (Fig. 2(b)) are shown for different values of $k$. To calculate standard deviation $\sigma_l$ in Eq. (5) and local standard deviation $\sigma_l(m)$ in Eq. (6), the time average is taken over $t=4 \times 10^5$ time units after an initial transient of $10^5$  units. The existence of disordered, chimera, multi-chimera and coherent states are represented by SI and DM. As mentioned above, (SI, DM)=(1, 0) represents a disordered state, while (SI, DM)=(0, 0) represents a coherent state. Further $0<\mbox{SI}<1$, DM=1 and $0<\mbox{SI}<1$, $2\leq \mbox{DM} \leq M/2$ represent chimera and multi-chimera states, respectively \cite{lakshmanan_m}.   For $r=0.3$, at lower values of $k$, we observe that all the neurons in the networks are in a disordered state, represented by the region I=$\{k: 0\leq  k\leq 0.72\}$. With an increase in the value of $k$ beyond $k=0.72$ we observe chimera and multi-chimera states in the region II=$\{k: 0.72 <  k < 1.24\}$. Chimera states represented by the value of $\mbox{DM}=1$ and multi-chimera states with $\mbox{DM}=2, 3, 4, 5, 6$ are shown in Fig. 2(b). In this state, all the neurons, namely coherent and disordered groups are in bursting type. With a further increase in the coupling strength we observe that all the neurons of the network are to be in coherent states represented by the region III=$\{k:1.24\leq  k \leq 2.0 \}.$ For $k>2.0$, all the neurons are always in a coherent state. Similarly, the variations of SI and DM for different values of the synaptic coupling strength $k$ are shown in Fig. 2(c) and Fig. 2(d), respectively, for $r=0.4.$ From Fig. 2, we observe that the extent of the region II for chimera and multi-chimera states gets decreased with increasing value of the coupling radius $r$.  Fig. 3 shows the two-parameter $(r, k)$ phase diagram of coherent, chimera / multi-chimera and incoherent states. The range of synaptic coupling strength $k$ for chimera or multi-chimera states is large for small values of coupling radius $r$, that is the network is almost locally coupled. But this region decreases for higher values of $r$ near to 0.5, that is when the system is globally coupled. Finally we have confirmed the above results for larger sizes of the network, namely N = 300 and N = 500 neurons to make sure that our results hold good for larger networks as well.

\begin{figure}[ht]
  \centering
  \includegraphics[width=0.52\textwidth]{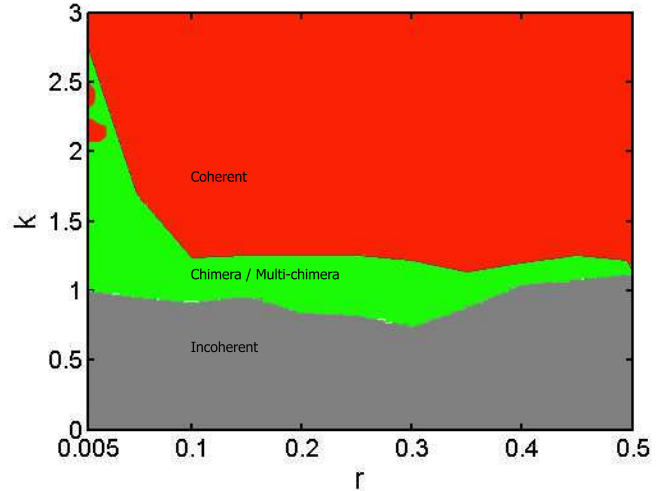} \\[0.5cm]
  \caption{(Color online) Two parameter $(r, k)$ phase diagram for $N=200$ nonlocally coupled network of identical Hindmarsh-Rose oscillators. Strength of incoherence is used as a measure for incoherence, coherence and chimera / multi-chimera states. Red region is for coherent, green is for chimera or multi-chimera and gray is for incoherent states. }
  \label{Fig1}
\end{figure}

\section{Global interaction}
\par We can convert the non-local interaction into a global interaction by taking the number of nearest neighbors as $p=\frac{N-1}{2}$, where the number of nodes $N$ in the network is an odd number. Then the network of identical Hindmarsh-Rose oscillators with global coupling is as follows:
$$\dot{x_i}=ax_{i}^{2}-x_{i}^{3}-y_{i}-z_{i}+\frac{k}{N-1}(v_{s}-x_{i})\sum_{j=1}^{N} c_{ij} \Gamma (x_{j}), \hskip 70pt$$
$$\dot{y_i}=(a+\alpha )x_{i}^{2}-y_{i}, \hskip 160pt \eqno{(9)}$$
$$\dot{z_i}=c(b~x_{i}-z_{i}+e), \;\; i=1, 2, \cdot \cdot \cdot , N \hskip 90pt$$
where $C=(c_{ij})_{n \times n}$ is the connectivity matrix  such that $c_{ij}=1$ if $i \not = j$ and $c_{ii}=0.$ 
\par In order to understand the existence of mutual synchronization of the above globally coupled network, we carry out an analytical investigation based on a linear stability analysis, closely following the stability analysis for a synchronized manifold by Belykh et al. \cite{Igor, Igor_2015}.
  For the synchronized state $x_i(t)=x(t), y_i(t)=y(t), z_i(t)=z(t)$ for all $i$, and so the system (9) becomes \\
$$\dot{x}= ax^{2}-x^{3}-y-z+\frac{Rk}{N-1}(v_{s}-x)\Gamma (x), \hskip 80pt$$
$$\dot{y}=(a+\alpha)x^{2}-y, \hskip 170pt \eqno{(10)}$$
$$\dot{z}=c(b~ x-z+e),\hskip 190pt$$ 
where $R$ is the row sum of the connectivity matrix $C$. Introducing the differences between the neural oscillator coordinates, 
$$\xi_{ij}=x_{j}-x_{i}, \eta_{ij}=y_{j}-y_{i}, \zeta_{ij}=z_{j}-z_{i}, i,j=1,2,\cdot \cdot \cdot ,N,  \eqno{(11)}$$
\\the linearized stability equations for the transverse perturbation of the synchronization manifold are given by\\

$$\dot{\xi}_{ij}=(2ax-3x^{2})\xi_{ij}-\eta_{ij}-\zeta_{ij}-\frac{Rk\Gamma(x)}{N-1}\xi_{ij} \hskip 120pt$$
$$+\frac{k}{N-1}(v_{s}-x)\Gamma^{'}_{x}(x)(R\xi_{ij}+\sum_{h=1}^N(c_{jh}\xi_{jh}-c_{ih}\xi_{ih})),$$
$$\dot{\eta}_{ij}=2(a+\alpha)x~\xi_{ij}-\eta_{ij}, \hskip 170pt \eqno{(12)}$$
$$\dot{\zeta}_{ij}=c(b ~\xi_{ij}-\zeta_{ij}). \hskip 170pt$$   
where $x(t)$ is the synchronous solution defined by the system (10) and $\Gamma^{'}_{x}(x)$ is the partial derivative of $\Gamma(x)$ with respect to $x$. The term $\sum_{h=1}^N(c_{jh}\xi_{jh}-c_{ih}\xi_{ih})$ is the same as in the case of linear coupling \cite{Igor, Pecora}. The derivatives are calculated at $\xi=0, \eta=0, \zeta=0$. Then the stability equations become \\
$$\dot{\xi}=(2ax-3x^{2})\xi-\eta-\zeta-\Omega(x)\xi, \hskip 170pt$$
$$\dot{\eta}=2(a+\alpha)x ~\xi-\eta, \hskip 170pt \eqno{(13)}$$
$$\dot{\zeta}=c(b ~\xi-\zeta), \hskip 190pt$$ 
where $\Omega(x)=\frac{Rk\Gamma(x)}{N-1}-\frac{k}{N-1}(v_s-x)\Gamma'_x(x)(R+\lambda_2),$ $\lambda_2$ is the largest real part of the eigenvalues of the coupling matrix $M=C-R I.$ It is well known that the matrix $M$ has one zero eigenvalue $\lambda_1$ and all the other eigenvalues have non-positive real parts \cite{Igor}. If the coupling is mutual then the coupling matrix $M$ is symmetric and all the eigenvalues are real. For simplicity, let us suppose that the largest eigenvalue $\lambda_2$ of the coupling matrix $M$ is simple. Equation (13) is then an analog of the Master Stability equation \cite{Pecora}. 
\begin{figure}[ht]
  \centering
  \includegraphics[width=0.52\textwidth]{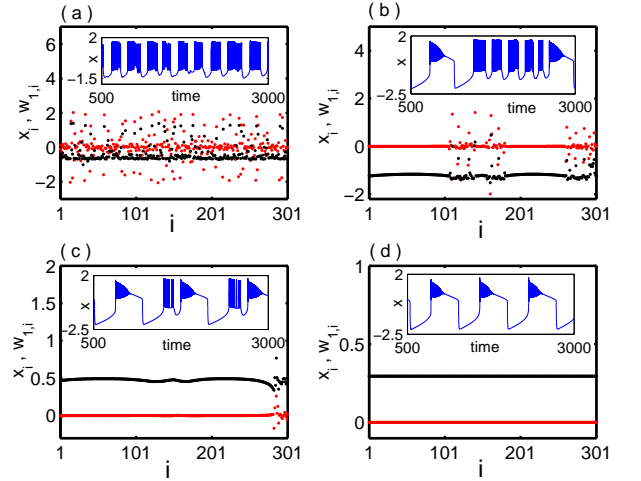} \\[0.6cm]
  \caption{(Color online) Snapshots of a system of globally coupled Hindmarsh-Rose neurons for different values of the synaptic coupling strength $k$ in terms of the variables $x_i$ (black color) and the transformed variables $w_{1, i}=x_i-x_{i+1}$ (red color): (a) incoherent state, $k=1.0$, (b) chimera state (with two synchronized/desynchronized groups), which can be reindexed into single group each, $k=1.2$, (c) chimera state (with single synchronized group), $k=1.28$, (d) coherent state, $k=1.3$. The inset figures are the corresponding time series (blue color). The number of oscillators is $N=301.$}
  \label{Fig1}
\end{figure}

\par Numerical results of the global synaptically coupled Hindmarsh-Rose neurons are presented in Fig. 4 for $N=301$. The initial conditions are chosen as follows: $x_{i0}=0.01(i-\frac{N-1}{2}), y_{i0}=0.02(i-\frac{N-1}{2}), z_{i0}=0.03(i-\frac{N-1}{2})$ for $i=1, 2, \cdot \cdot \cdot , \frac{N-1}{2}$ and $x_{i0}=0.1(\frac{N-1}{2}-i), y_{i0}=0.12(\frac{N-1}{2}-i), z_{i0}=0.21(\frac{N-1}{2}-i)$ for $i=\frac{N-1}{2}+1, \cdot  \cdot  \cdot , N$ with added small random fluctuations. The snapshot of the state variables $x_i$ and the new transformed variables $w_{1, i}=x_i-x_{i+1}$ are shown by black and red color dotted points, respectively. At a lower value of the synaptic coupling strength $k=1.0,$ all the neurons are in a disordered state (Fig. 4(a)) and their behavior are all of square-wave bursting in nature. At a higher value of $k=1.2,$ a typical pattern of the chimera state with two groups of synchronized/desynchronized oscillators, which of course can always be brought into separate single group of synchronized/desynchronized oscillators by appropriate reindexing of the oscillators (due to the global nature of the coupling), is observed between the neurons (Fig. 4(b)).  The behavior of a neuron in the coherent group and a neuron in the disordered group have the same time series form,  that is coexistence of square-wave and plateau bursting is observed in all the neurons. The typical time series of $x_i$ (blue color line) is shown in the inset of Fig. 4(b). The values of the transformed variables $w_{1, i}$ for the coherent groups are near zero whereas for the disordered neurons they are randomly distributed in [-2, 2]. With a further increase in the value of the synaptic coupling strength to $k=1.28$, chimera state with single coherent and incoherent group each is observed in which the left group of neurons are in coherence where the transformed variables take values close to zero and the right one is an incoherent one (Fig. 4(c)). All the coherent and disordered neurons exhibit a similar behavior, that is a mixture of square-wave and plateau bursting states shown in the inset of Fig. 4(c) by blue color line. At a further higher value of the synaptic coupling strength at $k=1.3$, all the neurons are found to be in a coherent or completely synchronized state (Fig. 4(d)). In this case all the neurons are in plateau bursting states. In globally coupled neurons, the region of chimera states are very small compared to non-local coupling, as already shown in Fig. 3 for $r=0.5$ and $N=200$.
\begin{figure}[ht]
  \centering
  \includegraphics[width=0.5\textwidth]{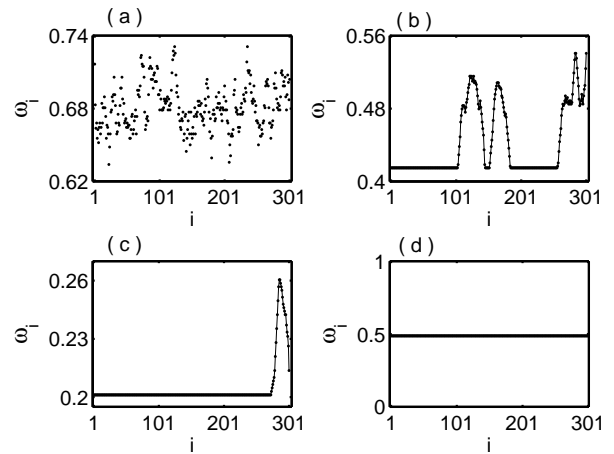} \\[-0.4cm]
  \caption{Mean phase velocities $\omega_i$ of globally coupled neurons corresponding to Fig. 4: (a) disordered state at $k=1.0$, (b) chimera state (with two synchronized / desynchronized groups) at $k=1.2$, (c) chimera state (with single synchronized group) at $k=1.28$, and (d) coherent state at $k=1.3$. Other parameters are same as in Fig. 4.}
  \label{Fig1}
\end{figure}

\par To confirm chimera states in globally coupled neurons, we calculate the mean phase velocity $\omega_i (i=1, 2, 3, \cdot \cdot \cdot , N)$ of each neuron. Figure 5 shows the mean phase velocity of each neuron corresponding to incoherent, chimera (with two synchronized / desynchronized groups), chimera (with single synchronized group) and coherent states. We have numerically integrated the dynamical equations (9) of the globally coupled neurons and the initial conditions used are the same as above. The time interval is taken over $t=5 \times 10^5$ time units  after an initial transient of $10^5$ units. We obtained similar results for even larger number of neurons, namely N=401 and 501, which confirm our above conclusions.

\begin{figure}[ht]
  \centering
  \includegraphics[width=0.5\textwidth]{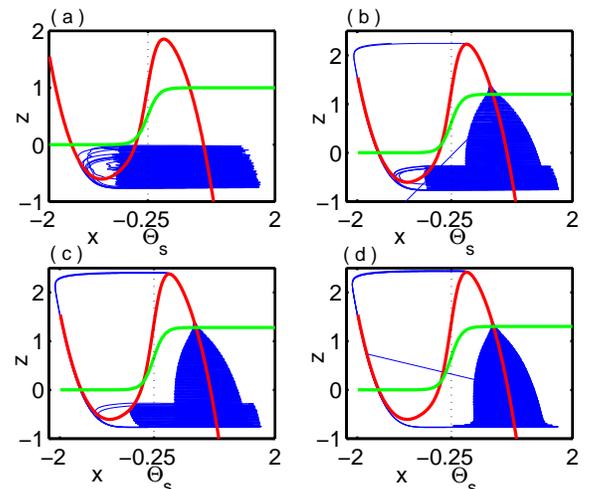} \\[-0.4cm]
  \caption{(Color online) Variation of stability function $\Omega(x)$ (green color), nullcline $z=f(x)$ (red color) and phase trajectory projected in $x-z$ plane (blue color) are shown for (a) disordered state at $k=1.0$, (b) multi-chimera state at $k=1.2$, (c) chimera state at $k=1.28$, and (d) coherent state at $k=1.3$.}
  \label{Fig1}
\end{figure}

\par The stability function for mutual synchronization is $\Omega(x)=\frac{Rk\Gamma(x)}{N-1}-\frac{k}{N-1}(v_s-x)\Gamma'_x(x)(R+\lambda_2),$ with $R=N-1$ and $\lambda_2=-N.$ It is strongly dependent on the membrane potential $x(t)$. The synaptic coupling strength $k$ is the upper bound of  $\Omega(x)$ for $x(t)\geq \Theta_s$ and rapidly decreases to zero if $x(t)< \Theta_s$. The global coupling yields a nullcline of $x$ as $z=f(x)=-\alpha x^2-x^3+\frac{Rk}{N-1}(v_s-x)\Gamma(x)$. The behavior of the stability function $\Omega(x)$ and nullcline $z=f(x)$ at incoherent, chimera (with two synchronized / desynchronized groups), chimera (with single synchronized group) and coherent states are shown in Figs. 6(a - d). It is important to note that at lower values of synaptic coupling strength, square-wave bursting is observed which turns into plateau bursting states for higher values of $k$ when coherent states occur in the globally coupled excitatory network (9). This happens when the synaptic coupling strength is large enough to change the square-wave to plateau bursting through a homoclinic bifurcation. By changing the synaptic coupling strength the system (9) undergoes transition from disordered to coherent states  corresponding to  the transition from square-wave to plateau bursting state via the disappearance of the homoclinic bifurcation. During the transition regime chimera states are observed. In the chimera states a combined bursting of square-wave and plateau is observed.


\section{Local interaction}
\par Next we consider a network of identical Hindmarsh-Rose oscillators with local (nearest neighbor ) interaction, that is $p=1$ in the non-local interaction (3), as follows:
$$\dot{x_i}=ax_{i}^{2}-x_{i}^{3}-y_{i}-z_{i}+\frac{k}{2}(v_{s}-x_{i})(\Gamma (x_{i-1})+\Gamma (x_{i+1})), \hskip 100pt$$
$$\dot{y_i}=(a+\alpha )x_{i}^{2}-y_{i}, \hskip 170pt \eqno{(14)}$$
$$\dot{z_i}= c(b~x_{i}-z_{i}+e), \;\; i=1, 2, \cdot \cdot \cdot , N, \hskip 200pt$$
with periodic boundary conditions  $(x_{0}=x_N, x_{N+1}=x_{1})$. As before, here $k$ is the synaptic coupling strength. 
\par A sequence of interesting behaviors occur numerically by changing the synaptic coupling strength $k$. We used the same initial conditions as was used in the case of non-locally coupled systems earlier in Sec. III. Fig. 7 shows the snapshots of the collective behavior of neurons for different values of the coupling strength $k$. We observe that at lower values of the coupling strength, all the neurons are in disordered states. This is shown in Fig. 7(a) for $k=0.4$. With increasing value of $k$ and beyond a critical value, we observe multi-chimera state in Fig. 7(b) for $k=1.2$. With further increase of $k$, this multi-chimera state is transformed into a chimera state in Fig. 7(c) for $k=1.36.$ For a further increase in the coupling strength to $k=3.6$ all the neurons are in a coherent state [Fig. 7(d)].
 \begin{figure}[ht]
  \centering
  \includegraphics[width=0.5\textwidth]{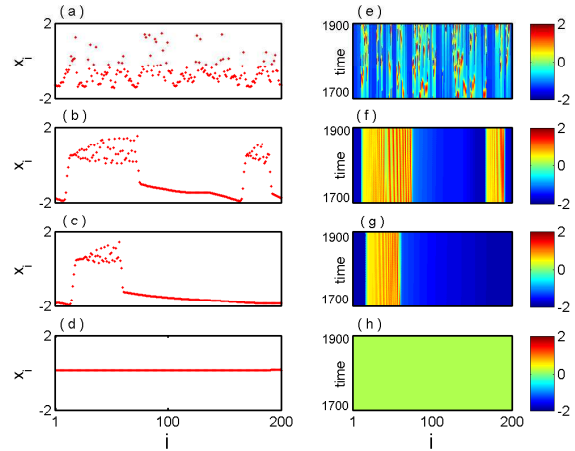} \\[-0.4cm]
  \caption{(Color online) Snapshots of amplitude (left panel) for (a) disordered state at $k=0.4$, (b) multi-chimera state at $k=1.2$, (c) chimera state at $k=1.36$, and (d) coherent state at $k=3.6$. Figs. (e)-(h) show spatio-temporal color coded maps for (a)-(d), respectively.}
  \label{Fig1}
\end{figure}
\begin{figure}[ht]
  \centering
  \includegraphics[width=0.5\textwidth]{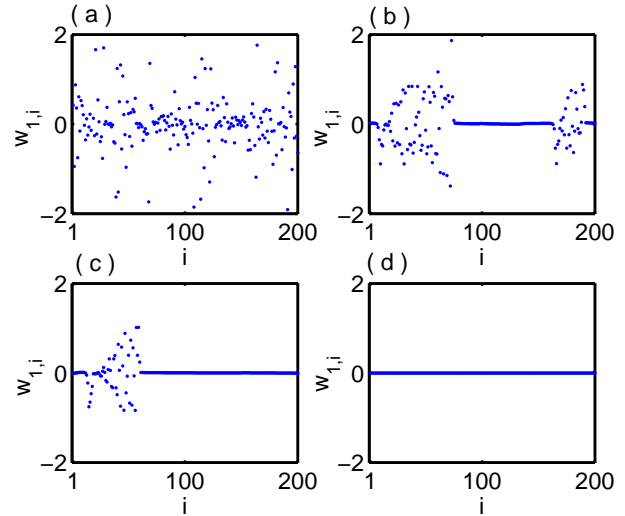} \\[-0.4cm]
  \caption{(Color online) Snapshots of the transformed variables $w_{1, i}$ corresponding to Fig. 7. (a) disordered state, $k=0.4$, (b) multi-chimera state, $k=1.2$, (c) chimera state, $k=1.36$ and (d) coherent state, $k=3.6$.}
  \label{Fig1}
\end{figure}

\par To further clarify the collective behavior, we plot the spatio-temporal nature
 of all the nodes $x_i(t)$ in the right panel of Fig. 7. Color coding of Fig. 7(e), 7(f), 7(g) and 7(h) clearly shows the presence of disordered, multi-chimera, chimera and coherent states for four distinct values of
 the synaptic coupling strength $k=0.4, 1.2, 1.36$, and 3.6, respectively. Snapshots of the locally coupled neurons (14) for different values of the synaptic coupling strength in terms of the new state variables $w_{1, i}=x_i-x_{i+1}$ are shown in Fig. 8. From this figure it is seen that initially at a lower value of the synaptic coupling strength, $k=0.4,$ all the neurons are in a disordered state and the values of $w_{1, i}$ are randomly distributed over [-2, 2] in Fig. 8(a). On increasing the synaptic coupling strength to $k=1.2$ [Fig. 8(b)] and $k=1.36$ [Fig. 8(c)], the values of $w_{1, i}$ for coherent group of neurons are near zero but the disordered group of neurons are randomly distributed over [-2, 2]. On further increasing the coupling strength to $k=3.6,$ [Fig. 8(d)] all the neurons are in a coherent state and the values of $w_{1, i}$ are near zero for all times. 

\par To confirm disordered, multi-chimera, chimera and coherent states, we have first calculated the mean phase velocity $\omega_i$ for each neuron which are shown in Fig. 9. The mean phase velocity $\omega_i$ for each neuron in disordered, multi-chimera, chimera and coherent states for $k=0.4, 1.2, 1.36$ and $3.6$ are shown in Figs. 9(a)-9(d) respectively. The values of mean phase velocity $\omega_i$ in Fig. 9(d) for each neuron is near to zero for the coherent state ($k=3.6).$ In the coherent state the behavior of all the neurons is close to the steady state of the locally coupled system $\dot{x}= ax^{2}-x^{3}-y-z+k(v_{s}-x)\Gamma (x), \dot{y}=(a+\alpha)x^{2}-y, \dot{z}=c(b~ x-z+e).$


\begin{figure}[ht]
  \centering
  \includegraphics[width=0.5\textwidth]{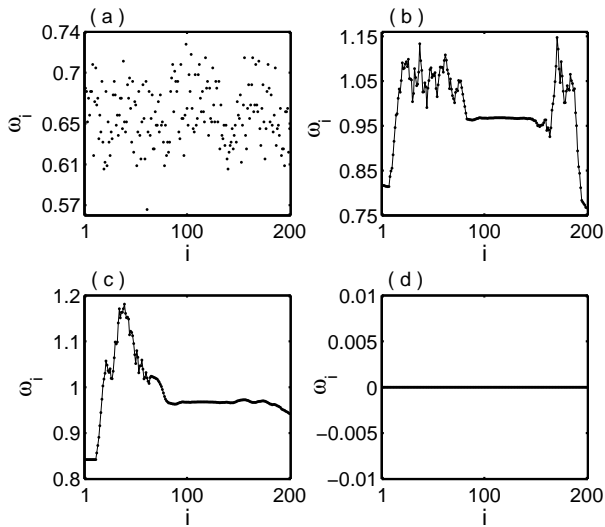} \\[-0.4cm]
  \caption{Values of mean phase velocity $\omega_i$ for each neuron corresponding to Fig. 7: (a) disordered state, $k=0.4$, (b) multi-chimera state, $k=1.2$, (c) chimera state, $k=1.36$ and (d) coherent state, $k=3.6$. Other value of parameters are same as Fig. 7.}
  \label{Fig1}
\end{figure}

In order to distinguish multi-chimera and chimera states clearly we use the strength of incoherence and discontinuity measure for different values of synaptic coupling strength $k.$ In Figs. 10(a) and 10(b) we demonstrate the behaviors of the strength of incoherence and discontinuity measure as a function of the synaptic coupling strength $k$ which clearly confirms the presence of chimera and multi-chimera states in the case of local interaction also. This appears to be quite surprising in the sense that chimera states have normally been identified only in the case of nonlocally or globally coupled arrays. Existence of chimera states using local coupling has been observed for larger number of neurons also, namely $N=(300, 500)$.

\begin{figure}[ht]
  \centering
  \includegraphics[width=0.52\textwidth]{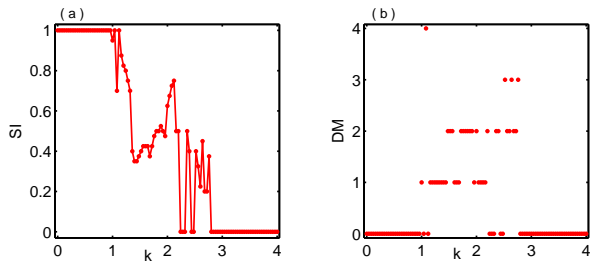} \\[-0.4cm]
  \caption{(Color online) Variation of (a) strength of incoherence (SI) and (b) discontinuity measure (DM) for different values of synaptic coupling strength $k$ for locally coupled Hindmarsh-Rose neurons. Here $N=200, M=40$ and $\delta=0.05.$ The time average $t=5000$ is considered after an initial transient of $10^5$  time units to calculate $\sigma_l$ and $\sigma_l(m)$ in Eqs. (5) and (6). }
  \label{Fig1}
\end{figure}

\par We also note that for local interaction in Eq. (3) $c_{ij}=1$ for either $j=i+1$ or $j=i-1$ so that (14) is valid in this case. For all other values of $j$, $c_{ij}=0$ and $c_{ii}=0.$  The value of eigenvalue $\lambda_2$ for a ring of $2K$-nearest neighbor mutually coupled neurons is $\lambda_{2}=-4\sum_{l=1}^K \mbox{sin}^{2}\frac{l\pi}{N}$ \cite{eigenvalue}.  In the case of local (nearest neighbor) interaction (Eq. (14)), $K=1$ and so the value of $\lambda_2$ is $\lambda_2=-4~ \mbox{sin}^2(\frac{\pi}{N})$ with $N=200$.\\
Now considering the master stability function \cite{Pecora} for the local synaptically coupled network (14), the stability function is   $\Omega(x)=Rk\Gamma(x)-k(v_{s}-x)\Gamma^{'}_{x}(x)(2+\lambda_{2})$ with $R=2$. For local excitatory coupling the nullcline of $x$ is $z=f(x)=-\alpha~ x^{2}-x^{3}+\frac{Rk}{2}(v_{s}-x)\Gamma(x)$. 
\begin{figure}[ht]
  \centering
  \includegraphics[width=0.5\textwidth]{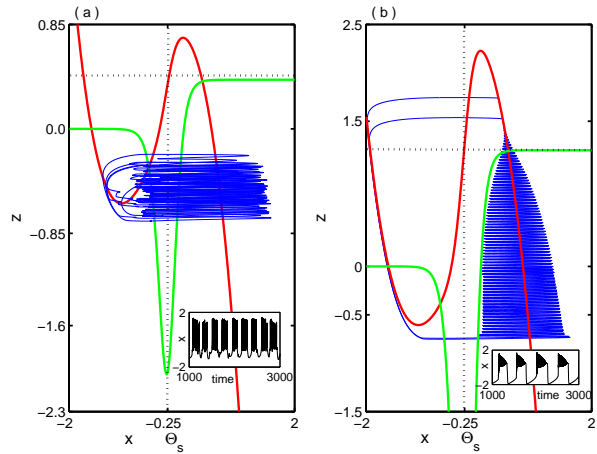} \\[-0.4cm]
  \caption{(Color online) The function $\Omega(x)$ (green color  line), nullcline $f(x)$ (red color line) and corresponding projected phase trajectories
 in $x-z$ plane (blue color line). (a) Incoherent bursting for $k=0.4$ and (b) multi-chimera state for $k=1.2$. The behaviour of the neuron (black color line) at $k=0.4$ and $k=1.2$ are shown in inset figures of (a) and (b) respectively.}
  \label{Fig1}
\end{figure}
\par The function $\Omega(x)$ corresponds to the contribution of the synaptic coupling for the stability of the synchronization manifold.
 In fact, in the case of global coupling the stability function $\Omega(x)$ varies between zero and synaptic coupling strength $k$ for all $x(t)$. On the other hand  for local coupling $\Omega(x)\leq 0$ for all values of the membrane potential $x(t)\leq \Theta_s$ and, while it has an
 upper bound $k$ for $x(t)>\Theta_s$ \cite{Igor, Igor_2015}. The first term of $\Omega(x)$ contains the Heaviside-type synaptic function $\Gamma(x)$ and it becomes significant when $x(t)\ge \Theta_s.$ The second term of $\Omega(x)$ is decisive for the values of $x$ near the threshold $\Theta_s$ as the derivative $\Gamma_x$ has a rapidly decaying tail. Practically the cells are uncoupled when the value of $x(t)$ is below the threshold $\Theta_s$.
\par At a lower value of the synaptic coupling strength $k=0.4,$ $\Omega(x)$ has a negative drop near $x=\Theta_s=-0.25$ (Fig. 11(a)) which signifies the incoherent state of the bursting neuron. The corresponding time series of the bursting neuron is shown in the inset of Fig. 11(a). With a further increase of coupling at $k=1.2,$ the value of $\Omega(x)$ is negative and has a sharp drop near $x=\Theta_s.$ The neurons are in a multi-chimera state. The coherent and disordered neurons are in a plateau bursting state in Fig. 11(b) which is in contrast to the chimera states studied earlier in the literature
 in chaotic systems, where the coherent states are in periodic states or remain close to steady states and disordered states are in chaotic states \cite{chaotic_chimera}. But in a chimera or multi-chimera state in the present case all the coherent as well as the  disordered neurons are in plateau bursting states.

\section{Conclusion}
In conclusion, we have analyzed the role of the chemical synaptic coupling function in inducing chimera and multi-chimera states of bursting neurons using non-local, global and local (nearest neighbor) interaction of neurons. Surprisingly, we find that chimera or multi-chimera states occur even in the presence of
 local interaction alone whereas previous studies \cite{strogatz,all_chimera} of chimera or multi-chimera states exist using either non-local or global interaction only. Interestingly, we identified three types of chimera states using different types of coupling configurations. In the first type, in the case of non-local coupling, in chimera or multi-chimera states, coherent and disordered neurons are all of bursting type. 
In the second type, we find that in the chimera or multi-chimera state using global synaptic coupling both the groups, namely the coherent or disordered groups, are in a combination of square-wave and plateau bursting states. That is for some times they are square-wave bursting in nature and for other times they are in a plateau bursting state (as shown in Fig. 4(b) or 4(c)). In the third type of chimera state using local synaptic coupling, we identified that both the groups in chimera / multi-chimera states are in plateau bursting states (Fig. 11(b)). Using suitable statistical measures disordered, multi-chimera, chimera or coherent states are confirmed. The existence of chimera and coherent states are also confirmed by using mean phase velocities \cite{mpv_PRL_2013}. To conclude, we wish to point out that this work promises to identity the existence of chimera states in various types of coupling topologies in bursting cells. Some types of chimera states might be helpful for information processing in the case of neurological diseases. Existence of chimera states in two dimensional grid of oscillators is more realistic in neurobiology. In our manuscript we observed chimera state using nonlocal, global and local type of interactions. Since interaction in two dimensional grid of oscillators are also of these types, we believe that chimera states can also emerge in these cases too. We are presently exploring this phenomenon and hope to report the results in the near future. 
\par Finally, it will be also of interest
 to extend our work to identify chimera or multi-chimera states in other systems using nonlinear local coupling. Moreover, the question of introducing
 time-delay in the nonlinear local coupling for the existence of chimera states \cite{bidesh} is also an important task for future studies.

{\bf Acknowledgments} M. L. has been supported by a DAE Raja Ramanna fellowship. We thank R. Gopal for the help in the numerical calculations of mean-phase velocity.

\end{document}